\documentclass[11pt]{article}
\usepackage[pctex32]{graphics}

\def\ben{\begin{equation}}
\def\een{\end{equation}}

\def\nn{\nonumber} \def\bd{\begin{document}} \def\ed{\end{document}}
\def\ds{\documentstyle} \let\fr=\frac \let\bl=\bigl \let\br=\bigr
\let\Br=\Bigr \let\Bl=\Bigl
\let\bm=\bibitem
\let\na=\nabla
\let\pa=\partial \let\ov=\overline
\newcommand{\be}{\begin{equation}}
\newcommand{\ee}{\end{equation}}
\def\ba{\begin{array}}
\def\ea{\end{array}}
\def\ft#1#2{{\textstyle{\frac{\scriptstyle #1}{\scriptstyle #2} } }}
\def\fft#1#2{{\frac{#1}{#2}}}
\def\del{\partial}
\def\vp{\varphi}
\def\sst#1{{\scriptscriptstyle #1}}
\def\oneone{\rlap 1\mkern4mu{\rm l}}
\def\td{\tilde}
\def\wtd{\widetilde}
\def\ie{{\it i.e.\ }}
\def\dalemb#1#2{{\vbox{\hrule height .#2pt
        \hbox{\vrule width.#2pt height#1pt \kern#1pt
                \vrule width.#2pt}
        \hrule height.#2pt}}}
\def\square{\mathord{\dalemb{6.8}{7}\hbox{\hskip1pt}}}
\newcommand{\ho}[1]{$\, ^{#1}$}
\newcommand{\hoch}[1]{$\, ^{#1}$}
\newcommand{\bea}{\setlength\arraycolsep{2pt} \begin{eqnarray}}
\newcommand{\eea}{\end{eqnarray}}
\newcommand{\ra}{\rightarrow}
\newcommand{\lra}{\longrightarrow}
\newcommand{\Lra}{\Leftrightarrow}
\newcommand{\bp}{\tilde \beta^\prime}
\newcommand{\tr}{{\rm tr} }
\newcommand{\Tr}{{\rm Tr} }
\def\0{{\sst{(0)}}}
\def\1{{\sst{(1)}}}
\def\2{{\sst{(2)}}}
\def\3{{\sst{(3)}}}
\def\4{{\sst{(4)}}}
\def\5{{\sst{(5)}}}
\def\6{{\sst{(6)}}}
\def\7{{\sst{(7)}}}
\def\8{{\sst{(8)}}}
\def\m{{\sst{(m)}}}
\def\n{{\sst{(n)}}}
\def\cA{{{\cal A}}}
\def\cB{{{\cal B}}}
\def\cF{{{\cal F}}}
\def\cG{{{\cal G}}}
\def\cH{{{\cal H}}}
\def\tV{\widetilde V}
\def\tW{\widetilde W}
\def\tH{\widetilde H}
\def\tE{\widetilde E}
\def\tF{\widetilde F}
\def\tA{\widetilde A}
\def\im{{{\rm i}}}
\def\tY{{{\wtd Y}}}
\def\ep{{\epsilon}}
\def\vep{{\varepsilon}}
\def\bD{{{\bar D}}}
\def\R{{{\mathbb R}}}
\def\C{{{\mathbb C}}}
\def\H{{{\mathbb H}}}
\def\CP{{{\mathbb C}{\mathbb P}}}
\def\RP{{{\mathbb R}{\mathbb P}}}
\def\Z{{{\mathbb Z}}}
\def\bA{{{\mathbb A}}}
\def\bB{{{\mathbb B}}}
\def\bC{{{\mathbb C}}}
\def\bD{{{\mathbb D}}}
\def\bE{{{\mathbb E}}}
\def\bZ{{{\mathbb Z}}}
\def\Re{{{\frak{Re}}}}
\def\Im{{{\frak{Im}}}}
\def\cosec{{\,\hbox{cosec}\,}}
\def\Gm{{\Gamma_{\!\! -}}}
\def\Gp{{\Gamma_{\!\! +}}}
\def\stan{{standard }}
\def\nonstan{{supernumerary }}
\def\p{{\partial}}
\def\kdel#1{{\fft{\del}{\del#1}}}

\def\bog{{Bogomolny }}
\def\om{{\hat{\omega}}}

\newcommand{\nnr}{\nonumber \\}
\newcommand{\pd}{\partial}
\newcommand{\ud}{\textrm{d}}
\newcommand{\dTH}{T^{\prime \, 0}_\textrm{H}}
\newcommand{\dOi}{\Omega^{\prime \, 0}_i}
\newcommand{\bx}{{\bf x}}

\begin{document}

\vspace{5mm}
\begin{center}
{\Large \bf Massive graviton propagation of the deformed Ho\v{r}ava-Lifshitz
gravity without projectability condition } \vspace{12mm}

 \centerline{\large Yun Soo Myung$^{a}$}

\vspace{10mm} {\em Institute of Basic Science and School of
Computer Aided Science \\ Inje University, Gimhae 621-749, Korea}
\vskip .6cm
\end{center}

\begin{center}

\underline{Abstract}
\end{center}

  We   study   graviton propagations of  scalar, vector,
  and tensor modes in the deformed Ho\v{r}ava-Lifshitz gravity ($\lambda R$-model) without projectability condition.
  The quadratic Lagrangian is invariant under diffeomorphism only for $\lambda=1$ case,
   which contradicts to the fact that $\lambda$ is irrelevant to a consistent Hamiltonian approach to the $\lambda R$ model.
    In this case, as far as scalar propagations are concerned,
  there is no essential difference between deformed Ho\v{r}ava-Lifshitz gravity ($\lambda R$-model) and general relativity.
    This implies that there are two degrees of freedom for a massless graviton without Ho\v{r}ava scalar, and five degrees of freedom appear for a massive graviton when  introducing Lorentz-violating  and Fierz-Pauli mass terms. Finally,
it is shown  that for $\lambda=1$, the vDVZ
discontinuity is absent in the massless limit of Lorentz-violating
mass terms by considering external source terms.

\vspace{15pt} \baselineskip=18pt
 \noindent $^a$ysmyung@inje.ac.kr

\thispagestyle{empty}

\newpage
\section{Introduction}
Recently Ho\v{r}ava has proposed a renormalizable theory of gravity
at a Lifshitz point~\cite{ho1},  which  may be regarded as a UV
complete candidate for general relativity.  Very recently, the
Ho\v{r}ava-Lifshitz gravity with a flow parameter $\lambda$ has been
intensively investigated
in~\cite{ho2,Vis,ho3,CHZ,KS,CNPS,SVW,KLM,BPS,Park,BS}. There are two
versions of Ho\v{r}ava-Lifshitz gravity in the literature:
 with/without the
projectability condition~\cite{muk}. Ho\v{r}ava has originally
proposed the projectability condition with/without the detailed
balance condition.   We  mention that the IR vacuum of this theory
is anti de Sitter (AdS) spacetimes. Hence, it is interesting to take
a limit of the theory, which leads  to  a Minkowski vacuum in the IR
limit.  To this end, one may modify the theory by including
``$\mu^4R$" and then, taking the $\Lambda_W \to 0$ limit~\cite{KS}.
This deformed Ho\v{r}ava-Lifshitz (dHL) gravity does not alter the
UV properties of the theory.  We note that the dHL gravity is
composed of $\lambda R$-model and higher spatial derivative terms
from detailed balance condition.  As far as the scalar propagations
are concerned, the essential part is the $\lambda R$-model because
most issues arose from this model.

Concerning the projectability condition, its role should be dealt
with carefully.  Actually, there exists a close relation between
projectability condition  and scalar degrees of freedom. The
projectability condition requires that the perturbation $A$  of the
lapse function $N$ depends only on time. It means that $A=A(t)$ is
not a Lagrange multiplier but a parameter. More seriously, by
imposing this condition at the beginning, one found  the global
Hamiltonian constraint instead of the local one. This  implies that
with  the projectability condition, the general relativity could not
be recovered from the dHL gravity with any $\lambda$.

 An urgent issue of the dHL
gravity is still to answer to the question of whether it can
 accommodate the Ho\v{r}ava scalar $\psi$,
  in addition to two physical degrees of freedom (DOF) for a massless graviton.
We would like to mention a few of relevant works. The
authors~\cite{CNPS} have shown that without the projectability
condition, the Ho\v{r}ava scalar $\psi$ is related to a scalar
degree of freedom appeared in  the massless limit of a massive
graviton.  Especially for the Hamiltonian approach to the
 dHL gravity, the authors~\cite{LP} did not consider the Hamiltonian
 constraint as a second class constraint, which leads to a  strange
 result that there are no DOF  left when  imposing the
 constraints of the theory. Moreover, the authors~\cite{HKG} have claimed that
 there are no solution of the lapse function which  satisfies  the
 constraints. Unfortunately, it implies  a surprising conclusion that there is no
 evolution  at all  for any observable. More recently, it
 was shown that the $\lambda
 R$-model (IR version of dHL gravity) which is considered as a gauge-fixed version of  general relativity  is  equivalent to the general relativity for any
 $\lambda$  when employing a consistent Hamiltonian formalism based on the Dirac algorithm~\cite{PT,BR}.
Although these has made a progress toward a consistent Hamiltonian
approach to the dHL gravity, there remains a subtle issue on the
equivalence\footnote{For example, one may find a vacuum torus
universe of $N=0$~\cite{Carl}  by assuming technical steps: First,
Eq.(20) in Ref.\cite{BR} is  multiplied by the lapse function $N$.
Then, integrating Eq.(20) over a whole space and finally, requiring
$R>0$. This is confirmed from the vacuum Hamiltonian constraint
Eq.(8) together with a second class constraint $\pi=0$: $R>0$ or
$\pi_{ij}=0$. This implies that there is no gravitational waves in
the torus universe, which seems contrary to the general relativity.
However, we have to admit that the torus universe is not an outcome
of the consideration, but it appears as a result of assuming
technical steps to avoid subtlety due to the boundary contribution.
We thank anonymous referee for pointing out this point. }

With the projectability condition, the authors~\cite{SVW,BPS} have
argued that $\psi$ is propagating around the Minkowski space but it
has a negative kinetic term, showing a ghost instability. In this
case, the Ho\v{r}ava scalar becomes ghost if the sound speed square
($c^2_\psi$) is positive. In order to avoid a ghost instability, the
sound speed square must be negative, but it is inevitably unstable
(gradient instability). Thus, one way to avoid this is to choose the
case that the sound speed square is close to zero ($c^2_\psi \to
0$), which implies the limit of $\lambda \to 1$. Unfortunately, in
the limit of $\lambda \to 1$, the cubic interactions are important
at very low energies~\cite{KA}. This invalidates any linearized
analysis and any predictability of quantum gravity is lost due to
unsuppressed loop corrections. This strong coupling problem appears
for an interacting theory of dHL gravity beyond the linearized
theory. This casts serious doubts on the UV completeness of the
theory.  Also, it was shown that  adding the mass term does not cure
a ghost instability in the Ho\v{r}ava scalar~\cite{Myungm}. However,
it was suggested that there are many ways to tame the gradient
instability of Ho\v{r}ava scalar~\cite{IM}. These are included (i)
the time scale is required to be longer than either the Jeans time
scale or the Hubble time scale (ii) higher spatial derivatives would
stabilize this instability when considering the dispersion relation
(iii) a phenomenological constraint on the renormalization group
flow may resolve the instability.

On the other hand, the authors~\cite{BPS2} have  tried to extend the
theory to make a healthy Ho\v{r}ava-Lifshitz gravity. However, there
has been some debate as to whether this  theory is really healthy.
The authors~\cite{PS} considered the IR limit of this theory and
showed that it suffered from the  strong coupling problem, too. To
response it, the original authors~\cite{BPS3} have claimed that the
strong coupling scale might exceed the cut-off scale for the
derivative expansions and thus, it seems to be  no strong coupling
issue. More recently, the authors~\cite{KP} has argued that the
alleged strong coupling problem is genuine and not merely an
artifact of a truncation the derivative expansion.

Hence, a current status of the dHL gravity may be summarized: the
projectability condition from condensed matter physics may not be
appropriate for describing the (quantum) gravity. Instead, if one
does not impose the projectability condition, the dHL gravity may
lead to general relativity without the strong coupling problem in
the IR limit.

  Inspired by a recent work of the consistency of the $\lambda R$-model (IR version of
dHL gravity)~\cite{BR}, we  will perform  a perturbation analysis of
the dHL
  gravity  without the projectability condition thoroughly.
  In this work, without the projectability condition, we investigate  massive graviton propagations of  scalar, vector, and tensor modes
 in the perturbation of dHL  gravity
  by introducing Lorentz-violating mass term (\ref{mass1}) and  Fierz-Pauli mass term (\ref{mass2}).
A motivation of the introduction  of these mass terms is to
investigate the strong coupling problem and the vDVZ discontinuity.
Even these mass terms violate the full diffeomorphism symmetry
without the projectability condition, it provides more DOF  through
the spontaneous symmetry breaking: less symmetry means more degrees
of freedom. Hence, we expect the change that 2 DOF (for massless
theory) $\to$ 5 DOF (for massive theory) including the Ho\v{r}ava
scalar.  We will show that the strong coupling problem is not
serious for vector and scalar modes when choosing Lorentz-violating
mass term~\cite{Rub}.  We will confirm that the Ho\v{r}ava scalar
survives in the massless limit of Fierz-Pauli mass term (vDVZ
discontinuity), but it is absent in the massless limit of
Lorentz-violating mass term (no vDVZ discontinuity)~\cite{CNPS}.

\section{dHL gravity}

First of all, we introduce the ADM formalism where the metric is
parameterized as
\be ds_{ADM}^2= - N^2  dt^2 + g_{ij} \Big(dx^i - N^i dt\Big)
\Big(dx^j - N^j dt\Big)\,, \ee
Then, the Einstein-Hilbert action can be expressed as
\be \label{Eins} S^{EH} = \fft{1}{16\pi G} \int d^4x \sqrt{g} N
\Big(K_{ij} K^{ij} - K^2 + R \Big)\,, \ee
where $G$ is Newton's constant and extrinsic curvature $K_{ij}$
takes the form
\be K_{ij} = \fft{1}{2N} \Big(\dot g_{ij} - \nabla_i N_j -
\nabla_j N_i\Big)\,. \ee
Here, a dot denotes a derivative with respect to $t$(
$``~\dot{}~"=\frac{\partial}{\partial t}$).

On the other hand, the  action of the dHL  gravity is given
by~\cite{KS}
\bea%
\label{act}S^{dHL}&=&\int dtd^3\bx\, \Big({\cal L}^0 +\sqrt{g}N\mu^4R + {\cal L}^h\Big)\,,\\
{\cal L}^0 &=& \sqrt{g}N\left\{\frac{2}{\kappa^2}(K_{ij}K^{ij}
\label{action1}-\lambda K^2)+\frac{\kappa^2\mu^2(\Lambda_W R
  -3\Lambda_W^2)}{8(1-3\lambda)}\right\}\,,\\ {\cal L}^h&=&
\sqrt{g}N\left\{\frac{\kappa^2\mu^2 (1-4\lambda)}{32(1-3\lambda)}R^2
-\frac{\kappa^2}{\eta^4} \left(C_{ij} -\frac{\mu
\eta^2}{2}R_{ij}\right) \left(C^{ij} -\frac{\mu
\eta^2}{2}R^{ij}\right) \right\}\,.\label{action2}
\eea%
Here $C_{ij}$ is the Cotton tensor defined by
\be C^{ij}=\epsilon^{ik\ell}\nabla_k\left(R^j{}_\ell
-\frac14R\delta_\ell^j\right)\label{def.K.C} \ee which is obtained
from the variation of gravitational Chern-Simons term with coupling
$1/\eta^2$.
 The full
equations of motion were derived in \cite{KK} and \cite{LMP}, but we
do not write  them  due to the length. Taking a limit of $\Lambda_W
\to 0$ in ${\cal L}^0+\sqrt{g}N\mu^4 R$, we obtain  the
  $\lambda R$-model ~\cite{KS}
 \be
S^{\lambda R}\equiv \int dt d^3x \tilde{\cal L}^{\lambda R}=\int dt
d^3x \sqrt{g} N\Bigg[\frac{2}{\kappa^2}\Big(K_{ij}K^{ij}-\lambda
K^2\Big)+\mu^4 R\Bigg]\ . \label{SM2} \ee  Comparing Eq.(\ref{SM2})
with general relativity (\ref{Eins}), the speed of light and
Newton's constant  are determined by
\be c^2=\fft{\kappa^2\mu^4}{2},~~
G=\fft{\kappa^2}{32\pi\,c},~~\lambda=1.\label{cg} \ee
Since we consider the $z=3$ Ho\v{r}ava-Lifshitz gravity, scaling
dimensions are  $[t]=-3,[x]=-1, [\kappa]=0,$ $[\mu]=1$, and $[c]=2$.
Even though the scaling dimensions are relevant to the UV
properties, these are also necessary to define the linearized theory
of $z=3$ Ho\v{r}ava-Lifshitz gravity consistently. The reason is
that we have to keep the same dimensions six for all terms, although
couplings of the kinetic term ($2/\kappa^2$) and the sixth order
derivatives ($\kappa^2/2\eta^4$) are dimensionless. In order to  see
the UV properties of power-counting renormalizability, it is better
to switch from the $c=1$ units to
 (\ref{cg}) units that impose the scaling dimensions. Switching back to $c=1$
units leads to the case~\cite{BR} \be S^{\lambda R}_{IR}=\mu^4\int
dt d^3x \sqrt{g} N\Bigg[\Big(K_{ij}K^{ij}-\lambda K^2\Big)+ R\Bigg]\
\label{SM2IR} \ee which   is   suitable for discussing the IR
properties (large distances) of strong coupling problem and vDVZ
discontinuity.

The deformed Lagrangian which is relevant to our study takes the
form~\cite{KS} \bea
\label{delag1} \tilde{{\cal L}}\equiv\tilde{\cal L}^{\lambda R}&+&{\cal L}^h  \\
\label{delag2}=\sqrt{g}N
\Bigg[\frac{2}{\kappa^2}\Big(K_{ij}K_{ij}-\lambda
K^2\Big)&+&\mu^4\Big(R
+\frac{1}{2\omega}\frac{4\lambda-1}{3\lambda-1}R^2-\frac{2}{\omega}R_{ij}R_{ij}\Big) \\
\label{delag3}&+&\frac{\kappa^2\mu}{2\eta^2}\epsilon^{ijk}R_{il}\nabla_jR^l_k
-\frac{\kappa^2}{2\eta^4}C_{ij}C_{ij}\Bigg] \eea where a
characterized parameter $\omega$ is given by \be \omega=\frac{16
\mu^2}{\kappa^2}=\frac{16\sqrt{2}c}{\kappa^3}. \ee   Actually, the
Lagrangian (\ref{delag2}) is enough to describe  scalar and vector
propagations because (\ref{delag3}) from the Cotton tensor
contributes to  tensor propagations only. For $\lambda=1$, taking
the limit of $\omega \to \infty$ while keeping  $c^2=1$ fixed  is
equivalent to recovering the Einstein gravity ($\lambda=1 R$-model).
Explicitly, this limit implies $\kappa^2 \to 0(\mu^4\sim \kappa^{-2}
\to \infty)$ which means that the kinetic term and curvature term
$\mu^4R$ dominate over all higher order curvature terms. The
deformed Lagrangian (\ref{delag1}) can be redefined to be \be
\tilde{\cal L}={\cal L}_K+{\cal L}_V, \ee where ${\cal L}_K({\cal
L}_V)$ denote the kinetic (potential) Lagrangian with (without)
temporal derivative terms.

 We wish to consider
perturbations of the metric around Minkowski spacetimes, which is a
solution to the full Lagrangian (\ref{delag1})
 \be \label{decom1}
g_{ij}= \delta_{ij}+\eta h_{ij},~N= 1+ \eta n,~N_i= \eta n_i, \ee
where a dimensionless coupling constant $\eta$ from gravitational
Chern-Simons term is included to define the perturbation. The
inclusion of $\eta$ makes sense because the non-interacting limit
corresponds to sending $\eta \to 0$, while keeping the ratio
$\gamma=\kappa/\eta$ fixed~\cite{ho1}. This in turn provides the
limit of $\kappa \to 0(\omega \to \infty)$.  For $\lambda=1$, this
limit yields a one-parameter family of free-field fixed points
parameterized by $\gamma$.

At quadratic order the $\lambda R$-action (\ref{SM2}) turns out to
be \bea \label{EHlambda} S^{\lambda R}_2 &=& \eta^2\int dt d^3x
\Bigg\{{1 \over \kappa^2} \left[{1\over 2} \dot h_{ij}^2
-{\lambda\over 2} \dot h^2 + (\partial_i n_j)^2 +(1-2\lambda)
(\partial \cdot n)^2 - 2
\partial_i n_j(\dot h_{ij} -\lambda \dot h \delta_{ij})\right]
\nonumber\\
&&\phantom{x} +  {\mu^4\over 2} \left[ -\frac{1}{2}(\partial_k
h_{ij})^2+\frac{1}{2}(\partial_i h)^2 +(\partial_i
h_{ij})^2-\partial_i h_{ij}\partial_j h + 2 n (\partial_i
\partial_j h_{ij}-\partial^2 h) \right]\Bigg\}  \eea
with $h=h_{ii}$.  A general Lorentz-violating mass term is given
by~\cite{Rub} \be \label{mass1}S_2^{LV}=\frac{\eta^2}{2\kappa^2}
\int dt d^3x \Big[ 4m_0^2
~n^2+2m_1^2~n_i^2-\tilde{m}_2^2~h_{ij}^2+\tilde{m}_3^2~
h^2+4\tilde{m}_4^2~ nh\Big]. \ee As was pointed out  in \cite{dub},
$S^{LV}_2$  provides  various phases of massive gravity in general
relativity.  In this work, we add  (\ref{mass1}) to the linearized
theory of dHL  gravity to investigate strong coupling problem and
the vDVZ discontinuity. In this work, we choose the case of $m_0=0$,
where the lapse field $n$ enters the action linearly and thus, it
still acts as a Lagrange multiplier. If one considers a non-zero
mass $m_0$ seriously, it induces a ghost instability~\cite{Rub,RT}.
At this stage, we would like to mention that for generic
backgrounds,  $m_1^2=0$ has provided  a well-defined case in
bi-gravity and massive gravity~\cite{BCNP,RT}. Also, the generic
case could be well behaved in generic backgrounds~\cite{BCNP}.

 We compare (\ref{mass1})  with the
Lorentz-invariant Fierz-Pauli mass term~\cite{FP}
 \be \label{mass2}
S_2^{FP}=\frac{ \eta^2}{2\kappa^2} \int dt  d^3x \Big[- m^2
h_{\mu\nu}h^{\mu\nu}+ m^2 \Big(h^\mu~_\mu \Big)^2\Big]. \ee

 In order to analyze  physical propagations thoroughly, it is convenient to use the cosmological
decomposition in terms of scalar, vector, and tensor modes under
spatial rotations $SO(3)$~\cite{MFB}
 \bea \label{pert}
 n &=&-\frac{1}{2}A,\nn \\
 n_i&=&\Big(\partial_iB+V_i\Big),\label{decom2} \\
 h_{ij}&=&\Big(\psi\delta_{ij}+\partial_i\partial_j E+2\partial_{(i}F_{j)}+t_{ij}\Big), \nn \eea
where the conditions of
$\partial^iF_i=\partial^iV_i=\partial^it_{ij}=t_{ii}=0$ are
imposed.
 The
last two conditions mean that $t_{ij}$ is a transverse and
traceless tensor in three spatial dimensions.  Using this
decomposition, the scalar modes ($A,B,\psi,E$), the vector modes
($V_i,F_i$), and the tensor modes ($t_{ij}$) decouple completely
from each other. These all amount to 10 degrees of freedom for a
symmetric tensor in four dimensions.

Before proceeding, let us check dimensions. Masses have scaling
dimensions: $[m_1^2]=2$ and
$[\tilde{m}_2^2]=[\tilde{m}_3^2]=[\tilde{m}_4^2]=6$. In order to get
the true mass with dimension 1, we redefine mass squares as \be
\tilde{m}_i^2=c^2 m^2_i, ~~{\rm for }~~ i=2,3,4 \ee which implies
that $ [m_i^2]=2$. The Fierz-Pauli mass term is recovered when all
masses  are equal except for $m_0$ as \be \label{mass3}
m_1^2=m_2^2=m_3^2=m_4^2=m^2;~m_0=0. \ee The quadratic  action for
$\lambda R$-model is obtained by substituting (\ref{decom2}) into
the quadratic action (\ref{EHlambda}) as
 \bea \label{fehl}
  S^{\lambda R}_2 =\frac{1}{2\gamma^2}
 \int dtd^3x &\Bigg\{&
  \Big[  3(1-3\lambda)\dot{\psi}^2
+2\partial_iw_j\partial^iw^j
-4\Big((1-3\lambda)\dot{\psi}+(1-\lambda)\partial^2\dot{E}\Big)\partial^2B
\nonumber\\
&&+4(1-\lambda)(\partial^2B)^2+2(1-3\lambda)\dot{\psi}\partial^2\dot{E}
+(1-\lambda)(\partial^2\dot{E})^2+ \dot{t}_{ij}\dot{t^{ij}}\Big]
\nonumber\\
&& +c^2\Big(2\partial_k\psi\partial^k\psi +4A\partial^2\psi
-\partial_k t_{ij}\partial^k t^{ij}\Big)\Bigg\} \eea with
$\gamma^2=\kappa^2/\eta^2$ and $w_i=V_i-\dot{F}_i$. We have  the
coupling of $\frac{1}{2\gamma^2}$ in  the quadratic action.
 The
higher order action  from ${\cal L}^h$ takes the form \be
S^h_2=\frac{\kappa^2\mu^2\eta^2}{8}\int dt d^3x \Bigg[
-\frac{1-\lambda}{2(1-3\lambda)} \psi
\partial^4 \psi -\frac{1}{4}t_{ij}\partial^4 t^{ij}
+\frac{1}{\mu \eta^2} \epsilon^{ijk} t_{il} \partial^4
\partial_j t^l~_k+\frac{1}{\mu^2\eta^4}
t_{ij} \partial^6 t^{ij}
 \Bigg]. \label{1action} \ee
We find  that  two modes of scalar $\psi$ and tensor $t_{ij}$ exist
in $S^h_2$  only, missing vector modes. Since the spatial slice is
conformally flat, the vanishing Cotton tensor and the absence of six
derivative terms  result in the scalar sector. Also, the Cotton
tensor does not contribute to vector modes ($V_i,\dot{F}_i$). The
vectors are absent in $S^h_2$  because the vector belongs to gauge
degrees of freedom in the massless gravity theory.

Before we proceed,  we mention  the foliation-preserving
diffeomorphism (FDiff) in the dHL gravity with the projectability
condition. Considering  the anisotropic scaling of temporal and
spatial coordinates ($t\to b^z t, x^i \to b x^i$), the time
coordinate $t$ plays a privileged role. A quadratic action of
$S^{\lambda R}_2+S^h_2$ should  be  invariant under FDiff whose
transformation is given by \be t \to \tilde{t}=t+\epsilon^0(t),~~x^i
\to \tilde{x}^i=x^i+\epsilon^i(t,{\bf x}), \ee which shows that  the
spacetime symmetry is smaller than the full diffeomorphism (Diff) in
the general relativity \be t \to \tilde{t}=t+\epsilon^0(t,{\bf
x}),~~x^i \to \tilde{x}^i=x^i+ \epsilon^i(t,{\bf x}). \ee  FDiff
(Diff) invariance are  dynamical symmetry of dHL gravity with the
projectability condition (general relativity) and not just symmetry
of  the background spacetimes. Hence, it controls the number of
propagating degrees of freedom: more symmetry means less degrees of
freedom. It is well known that general relativity  as a massless
gravity theory has two degrees of freedom, while the dHL gravity
with the projectability condition  may have three.

In this work, we consider the dHL gravity  without imposing  the
projectability condition. In this case, Diff is more suitable  than
FDiff.
  Using the notation of
$\epsilon^\mu=(\epsilon^0,\epsilon^i)$ and
$\epsilon_\nu=\eta_{\nu\mu}\epsilon^\mu$, the perturbation of metric
transforms as \be \delta g_{\mu\nu} \to
\delta\tilde{g}_{\mu\nu}=\delta g_{\mu\nu}+\partial_\mu
\epsilon_\nu+\partial_\nu \epsilon_\mu. \ee Further, making a
decomposition $\epsilon^i$ into a scalar $\xi$ and a pure vector
$\zeta^i$ as $\epsilon^i=\partial^i\xi+ \zeta^i$ with $\partial_i
\zeta^i=0$, one finds the transformation for scalars \be
\label{trans1} A \to \tilde{A}=A-2\dot{\epsilon^0},~\psi \to
\tilde{\psi}=\psi,~ B \to \tilde{B}=B-\epsilon^0+\dot{\xi},~E \to
\tilde{E}=E+2\xi.\ee On the other hand, the vector and the tensor
take the forms \be \label{trans2} V_i \to \tilde{V}_i=V_i
+\dot{\zeta}_i,~F_i\to \tilde{F}_i=F_i+\zeta_i,~t_{ij} \to
\tilde{t}_{ij}=t_{ij}. \ee  For the Diff transformations,  gauge
invariant combinations are \be t_{ij},~~w_i=V_i-\dot{F}_i, \ee for
tensor and vector, respectively and \be
\psi,~~\Phi=c^2A-\dot{\Pi}~~{\rm with}~~\Pi=2B-\dot{E} \ee for two
scalar modes\footnote{For TDiff respecting an  additional constraint
$\partial_\mu \epsilon^\mu=0$~\cite{ABGV}, there are three
gauge-invariant scalar modes: $\psi,~\Phi,$ and
$\Theta=A-\bigtriangleup E$. In this case, a truly propagating
scalar graviton is given by $\psi$.}. We note that $\Pi$ is not a
gauge-invariant scalar mode.

Let us  try to express the quadratic action (\ref{fehl}) in terms of
gauge-invariant quantities as~\cite{KLM} \bea S^{\lambda R}_2
=\frac{1}{2\gamma^2}
 \int dtd^3x &\Bigg\{&
\Big[  3(1-3\lambda)\dot{\psi}^2 -2w_i\bigtriangleup w^i
-2(1-3\lambda)\dot{\psi}\bigtriangleup \Pi+(1-\lambda)(\bigtriangleup\Pi)^2 \nonumber \\
\label{secondlam}&+& \dot{t}_{ij}\dot{t}^{ij}\Big]
+c^2\left(-2\psi\bigtriangleup\psi +4\bigtriangleup A\psi + t_{ij}
\bigtriangleup t^{ij}\right)\Bigg\} \eea with the spatial Laplacian
$\bigtriangleup=\partial^2$. However, it is no doubt  that for
general $\lambda$,  the quadratic action of $\lambda R$-model is not
expressed  in terms of gauge-invariant quantities. This contrasts to
the Hamiltonian approach which shows that the value of $\lambda$ is
completely irrelevant for finding two physical degrees of freedom
for a massless graviton~\cite{BR}. In the Hamiltonian approach, they
have used Diff as full dynamical symmetry and have chosen a
gauge-fixing to identify four degrees of freedom in phase space.
However, in this Lagrangian approach, we are working with  the
quadratic action  and not making a gauge-fixing. The Diff could be
manifestly realized at the quadratic action only for $\lambda=1$.
Then, the $\lambda=1$ case leads to the
 gauge-invariant action as
\be S^{\lambda=1 R}_2 =\frac{1}{2\gamma^2}
 \int dtd^3x \Bigg\{
\Big[  -6\dot{\psi}^2 -2w_i\bigtriangleup w^i +4\psi \bigtriangleup
\Phi+\dot{t}_{ij}\dot{t}^{ij}\Big] +c^2\Big[-2\psi\bigtriangleup\psi
 +t_{ij} \bigtriangleup t^{ij}\Big]\Bigg\}.
\ee At this stage, it is unclear why the value of $\lambda$ is
uniquely determined  to be 1 in the perturbation theory. Other
possibility includes the case that for generic $\lambda$, $\Pi$ and
$A$ are separately gauge-invariant scalars. However, this is not the
case. An allowable case is that for generic $\lambda$, $\Pi$ is a
gauge-invariant scalar and $A$ is a parameter, which is exactly the
dHL gravity with the projectability condition.    We note that
$S_2^h$ in (\ref{1action}) contains only $\psi$ and $t_{ij}$, which
are
 gauge-invariant quantities.

On the other hand, the mass term (\ref{mass1}) leads to
 \bea
\label{mass4}S_2^{LV}=\frac{1}{2\gamma^2} \int dt  d^3x \Bigg[
&2& m_1^2\Big(V_i^2+(\partial_iB)^2\Big) \nn\\
&-&\tilde{m}_2^2\Big(t_{ij}t^{ij}+2(\partial_iF_j)^2+(\partial_i\partial_j E)^2+2 \psi \partial^2 E+3 \psi^2\Big) \nn \\
&+&\tilde{m}_3^2\Big( \partial^2 E+3\psi\Big)^2-2\tilde{m}_4^2
A\Big(\partial^2 E+3\psi\Big) \Bigg]. \eea which is  not obviously
invariant under Diff  because we could not express whole terms in
terms of gauge-invariant quantities.  However, these do not give
rise to any problem because we are interested in the massless limit
of Lorentz-violating mass term and we do not impose  any gauge to
perform the perturbation analysis around the Minkowski background.

\section{Massive propagations}

Without the projectability condition, we conjecture that out of the
5 DOF of a massive graviton, 2 of these are expressed as transverse
and traceless tensor modes $t_{ij}$, 2 of these are expressed as
transverse vector modes $F_i$, and the remaining one is from
Ho\v{r}ava scalar $\psi$.

\subsection{Tensor modes} The field equation for tensors is given
by \be \label{tensormo} \ddot{t}_{ij}-c^2 \bigtriangleup t_{ij}
+c^2m^2_2 t_{ij}+\frac{2c^2}{\omega}\bigtriangleup^2t_{ij}-
\frac{\kappa^4\mu}{4\eta^2}\epsilon_{ilm}\partial^l\bigtriangleup^2t_j~^m-
\frac{\kappa^4}{4\eta^4} \bigtriangleup^3 t_{ij}=0. \ee The
requirement that these modes are not tachyonic gives the stability
condition \be m^2_2\ge 0. \ee In the absence of mass, these modes
describe the chiral primordial gravitational waves~\cite{BS,Myung4}.
These circularly polarized modes are possible because the Cotton
tensor $C_{ij}$ is present, making parity violation. In the presence
of mass term, it may describe massive chiral gravitational waves.

 \subsection{Vector modes} It is clear
from Eqs.(\ref{fehl}) and (\ref{mass4}) that $V_i$ enters the action
without temporal derivatives, that is, it is a non-dynamical field
in the massless theory.  A massive vector Lagrangian takes the form
\be \label{vector1} {\cal L}^v=\frac{1}{\gamma^2} \Bigg[
-w_i\bigtriangleup w^i+m^2_1 V_i^2-\tilde{m}_2^2 (\partial_iF_j)^2
\Bigg] \ee with $w_i=V_i-\dot{F}_i$. It is obvious that in the
absence of mass terms, $w_i$ is a non-propagating vector mode. We
integrate $V_i$ out using the field equation obtained by varying the
action with respect to $V_i$  \be \bigtriangleup
(V_i-\dot{F}_i)-m_1^2 V_i=0 \ee which implies \be
V_i=\frac{\bigtriangleup}{\bigtriangleup-m_1^2} \dot{F}_i. \ee
Plugging this expression into Eq.(\ref{vector1}) leads to be \be
\label{vector4} {\cal L}^v=\frac{1}{\gamma^2} \Bigg[
\frac{\bigtriangleup m_1^2}{\bigtriangleup -m_1^2} \dot{F}_i^2+
\tilde{m}_2^2 F_i \bigtriangleup F^i \Bigg]. \ee In order to obtain
a canonical action, we introduce a canonical vector field
$\tilde{F}_i$  defined by \be \label{cvf}
F_i=\frac{\gamma}{m_1}\sqrt{\frac{\bigtriangleup
-m_1^2}{2\bigtriangleup }} \tilde{F}_i \propto \frac{1}{m_1
M_{Pl}}\sqrt{\frac{\bigtriangleup -m_1^2}{2\bigtriangleup }}
\tilde{F}_i\ee in the  $c=1$  units.  Then, the Lagrangian
(\ref{vector4}) takes a canonical form  \be {\cal L}_c^v=\frac{1}{2}
\Bigg[ \dot{\tilde{F}}_i^2-\frac{m_2^2}{m_1^2}
(\partial_i\tilde{F}_j)^2-m_2^2 \tilde{F}_j^2 \Bigg]. \ee  Now let
us discuss the strong coupling issue. In order to discuss the strong
coupling problem, we first note that \be \frac{1}{8\pi
G}=\frac{4c}{\kappa^2}\equiv M^2_{Pl}, \ee which leads to a
 relation between $\gamma$ and Planck mass scale $M_{Pl}$
\be \label{imrel} \gamma=\frac{2\sqrt{c}}{\eta M_{Pl}} \propto
\frac{1}{M_{Pl}} \ee in the $c=1$ units.  Considering the relation
Eq.(\ref{cvf}), the original vector field is proportional to
$(mM_{Pl})^{-1}$ and from Eq.(\ref{vector1}), a gauge-invariant
combination $w_i$ takes the form \be w_i \propto \frac{m}{M_{Pl}}
\tilde{F}_i \ee which shows that vector modes at small $m$ is
precisely the same as in the Fierz-Pauli case.  The analysis in
Ref.\cite{AGS} suggests that  the strong coupling occurs at $E \sim
\sqrt{mM_{Pl}}$, which is a  high scale. In  comparison  to  the
Fierz-Pauli case, vector field changes nothing except the speed of
light.  Its equation of motion is given by \be
\ddot{\tilde{F}}_{i}-\frac{m_2^2}{m_1^2} \bigtriangleup
\tilde{F}_{j} +m^2_2 \tilde{F}_{i}=0, \ee which leads to the
dispersion relation \be \varpi^2=\frac{m^2_2}{m_1^2}k^2+m^2_2. \ee
For $m_1^2>0$ and $m_2^2>0$, it is obvious that  there is no ghosts.

In the Fierz-Pauli case of $m_1^2=m_2^2$, the massive vector
equation reduces to \be \Big(\square -m^2 \Big) \tilde{F}_{i}=0 \ee
which represents a massive vector with two degrees of freedom. Here
$\square=-\partial_0^2+\bigtriangleup$ with
$\partial_0=\frac{\partial}{\partial x^0}$ with $x^0=ct$.

\subsection{Scalar mode}

 The scalar Lagrangian with four different masses takes the form \bea \label{scalarL}
{\cal L}^s_\lambda =\frac{1}{2\gamma^2} &\Bigg[&-
3(3\lambda-1)\dot{\psi}^2 +2(3\lambda-1)\dot{\psi}\bigtriangleup
\Pi-(\lambda-1)(\bigtriangleup\Pi)^2
-\frac{\kappa^4\mu^2(1-\lambda)}{8(3\lambda-1)}\psi \bigtriangleup^2
\psi\nn
\\
&+&\mu^4\kappa^2 \Big(-\psi \bigtriangleup \psi+2 A\bigtriangleup
\psi\Big) -2m_1^2 B\bigtriangleup B
-\tilde{m}_2^2\Big(E\bigtriangleup^2E +2 \psi \bigtriangleup E +3
\psi^2\Big) \nn
\\
&+&\tilde{m}^2_3\Big(\bigtriangleup E+ 3\psi\Big)^2-2\tilde{m}^2_4
A\Big(\bigtriangleup E+3 \psi \Big)\Bigg]. \eea From the Lagrangian
(\ref{scalarL}), we find that there exist a Lagrange multiplier $A$
and a non-dynamical field $B$. Their variations with respect to $A$
are given by \be \label{scalar1} \bigtriangleup \psi -
\frac{\tilde{m}_4^2}{c^2}\Big(\bigtriangleup E+3\psi \Big)=0 \ee
which implies that $E$ can be expressed in terms of $\psi$  \be
\label{scalar2} E=\frac{2c^2
\psi}{\tilde{m}^2_4}-3\frac{\psi}{\bigtriangleup}. \ee  On the other
hand, the variation with respect to $B$ leads to \be \label{scalar3}
(3\lambda-1)\dot{\psi}+(\lambda-1)\bigtriangleup
\dot{E}-2(\lambda-1)\bigtriangleup B-m^2_1B=0, \ee Using
Eqs.(\ref{scalar2}) and (\ref{scalar3}), we can express $B$ in terms
of $\psi$ as \be \label{B-scalar}
B=\frac{2}{2(\lambda-1)\bigtriangleup+m_1^2}
\Bigg[(\lambda-1)\frac{2c^2\bigtriangleup
\dot{\psi}}{\tilde{m}_4^2}+ \dot{\psi} \Bigg]. \ee Hence, we rewrite
$\Pi$ in terms of Ho\v{r}ava scalar $\psi$ as \be \Pi=2B-\dot{E}
=\frac{4}{2(\lambda-1)\bigtriangleup+m^2_1}\Bigg[(\lambda-1)\frac{2c^2\bigtriangleup
\dot{\psi}}{\tilde{m}_4^2}+\dot{\psi}\Bigg]
-\frac{2c^2\dot{\psi}}{\tilde{m}_4^2}+\frac{3}{\bigtriangleup}\dot{\psi}.
\ee Plugging $E$, $B$, and $\Pi$ into (\ref{scalarL}) leads to a
very complicated Lagrangian for $\psi$ \bea  \label{compsl}{\cal
L}^\psi_{\lambda} =\frac{1}{2\gamma^2}
&\Bigg[&-3(3\lambda-1)\ddot{\psi} \psi
+4(3\lambda-1)\frac{\ddot{\psi}\bigtriangleup \psi }{m_4^2}
-\frac{\kappa^4\mu^2(1-\lambda)}{8(3\lambda-1)}\psi
\bigtriangleup^2 \psi \nn \\
&-&\frac{(3\lambda-1)}{2(\lambda-1)\bigtriangleup+m_1^2}\Bigg(8(\lambda-1)\frac{\ddot{\psi}
\bigtriangleup^2 \psi}{m_4^2}+8\ddot{\psi}
\bigtriangleup \psi \Bigg) \nn \\
&+&\frac{2(\lambda-1)}{2(\lambda-1)\bigtriangleup+m_1^2}\Bigg(-8(\lambda-1)\frac{\ddot{\psi}
\bigtriangleup^3
\psi}{m_4^4}+[12(\lambda-1)-8]\frac{\ddot{\psi}\bigtriangleup^2\psi}{m_4^2}
+12\ddot{\psi}\bigtriangleup \psi\Bigg) \nn \\
&+&\frac{16(\lambda-1)+\frac{8m_1^2}{\bigtriangleup}}{[2(\lambda-1)\bigtriangleup+m_1^2]^2}
\Bigg((\lambda-1)^2\frac{\ddot{\psi}\bigtriangleup^4 \psi}{m_4^4}+
2(\lambda-1)\frac{\ddot{\psi}\bigtriangleup^3\psi}{m_4^2}+\ddot{\psi}\bigtriangleup^2\psi\Bigg) \nn \\
&+&(\lambda-1)\Bigg(\frac{4\ddot{\psi}\bigtriangleup^2\psi}{m_4^4}-\frac{12\ddot{\psi}\bigtriangleup\psi}{m_4^2}
+9\ddot{\psi}
\psi\Bigg) \nn \\
&-&4\Bigg(\frac{m_2^2-m_3^2}{m_4^2}\Bigg)\psi (c^2\bigtriangleup^2)
\psi +\Bigg(8\frac{m^2_2}{m_4^2}-2\Bigg)\psi c^2\bigtriangleup \psi
-6c^2m_2^2 \psi^2 \Bigg]. \eea It seems difficult to derive the
equation of motion for the Ho\v{r}ava scalar. However, in the limit
of $\lambda \to 1$, we have a simplified Lagrangian for $\psi$ \bea
{\cal L}^\psi_{\lambda=1} =\frac{1}{2\gamma^2} \Bigg[-6 \ddot{\psi}
\psi &+&
\Big(\frac{8}{m_4^2}-\frac{8}{m_1^2}\Big)\ddot{\psi}\bigtriangleup
\psi-4\Bigg(\frac{m_2^2-m_3^2}{m_4^4}\Bigg) \psi c^2
\bigtriangleup^2 \psi \nn \\
\label{lam1psi}&+& \Bigg(8\frac{m^2_2}{m_4^2}-2\Bigg)\psi
c^2\bigtriangleup \psi -6c^2m_2^2 \psi^2 \Bigg], \eea which is
identified with the  quadratic Lagrangian for  general relativity
with the LV mass term~\cite{Rub}.  The equation of motion for $\psi$
is given by \bea
6\ddot{\psi}&-&8\Big(\frac{1}{m_4^2}-\frac{1}{m_1^2}\Big)\bigtriangleup
\ddot{\psi} \nn \\
\label{psiequation}&-&
2\Big(\frac{4m^2_2}{m_4^2}-1\Big)c^2\bigtriangleup \psi
+4c^2\Big(\frac{m_2^2-m_3^2}{m_4^4}\Big) \bigtriangleup^2 \psi
 +6c^2m_2^2 \psi=0. \eea
For $m_1^2>m_4^2,m_2^2>m_3^2, 4m^2_2>m_4^2$, there are no ghost
because all terms in the equation have correct signs
($\bigtriangleup$ is negative-definite). This contrasts to that of
dHL gravity with the projectability condition~\cite{Myungm}, which
indicates  that there is no ghost free, massive propagation for the
Ho\v{r}ava scalar $\psi$.

In order to discuss the strong coupling issue, let us remind the
relation of $\gamma \propto 1/M_{Pl}$ in the $c=1$ units.
Considering the Lagrangian (\ref{lam1psi}), we could define the
normalization factor relating $\psi$ and its canonically normalized
field $\psi^c$ as \be
\psi=\gamma\Bigg[-8\Bigg(\frac{1}{m_4^2}-\frac{1}{m_1^2}\Bigg)\bigtriangleup+6\Bigg]^{-1}
\psi^c \propto \frac{m}{M_{Pl}} \psi^c. \ee Also, considering
Eqs.(\ref{scalar2}) and (\ref{B-scalar}), one finds that  \be E,~B
\propto \frac{1}{mM_{Pl}}\psi^c \ee which are in a complete analogy
to vector modes. All these ensure that the strong coupling scale in
the scalar sector is the same as in vector sector ($E \sim \sqrt{m
M_{Pl}}$).

However,  in the case of Fierz-Pauli mass (\ref{mass3}), this
picture is changed. In this case, we have \be \psi=\gamma \psi^c \ee
and \be
 E,~B \propto
\frac{1}{m^2M_{Pl}}\psi^c \ee which implies the low energy scale of
strong coupling $(E \sim (m^4M_{Pl})^{1/5})$~\cite{AGS}.  Moreover,
the equation (\ref{psiequation}) reduces to a simpler equation \be
\label{fpequation} \ddot{\psi}-c^2\bigtriangleup \psi+c^2m^2 \psi=0
\ee  which is nothing but  the massive Klein-Gordon equation \be
\Big(\square -m^2 \Big) \psi=0. \ee

Finally, we confirm that a massive graviton takes five degrees of
freedom for both the Lorentz-violating  and Fierz-Pauli mass terms.

\section{vDVZ discontinuity}

Since the Lagrangian (\ref{compsl}) takes a complicated form, it is
a  formidable  task to prove that for generic $\lambda$, there is no
vDVZ discontinuity in the massless limit. Instead, we will show that
for $\lambda=1$ case, there is no the vDVZ discontinuity, in
comparison  to the Fierz-Pauli case.  In order to show it, we have
to introduce the external source term \be \label{extsou}
S_{int}=-\frac{1}{\gamma^2}\int dt
d^3x\Big[h^{ij}T_{ij}+2h^{0j}T_{0j}+h^{00}T_{00}\Big]. \ee
 A covariant form of the source conservation-law  $\partial_\mu T^{\mu\nu}=0$ is slightly
modified to  have \be \dot{T}_{00}=a\partial_j T_{j0},~~
\dot{T}_{0i}=\partial_jT_{ji} \ee where $a$  with $[a]=4$ is
inserted to have correct scaling dimensions
 \be
[T_{ij}]=6,~[T_{0j}]=4,~[T_{00}]=6. \ee On later, $a$ will be
determined to be  $a=c^2$. Then, we could express the above in terms
of gauge-invariant modes as \be \label{gisource}
S_{int}=-\frac{1}{\gamma^2}\int dt d^3x\Big[t_{ij}T_{ij}-2w_iT_{0i}+
\Big(a A-\dot{\Pi}\Big) \frac{T_{00}}{a}+\psi T_{ii}\Big]. \ee
Choosing $a=c^2$, we note that $c^2A-\dot{\Pi}$ is  nothing but a
gauge-invariant scalar $\Phi$  under Diff transformations.  We wish
to study  the vDVZ discontinuity by making use of (\ref{gisource}).

\subsection{$\lambda=1 R$-model}
First of all, we consider the quadratic action of $S_{2}^{\lambda=1
R}$ in (\ref{secondlam}) together with external source in
(\ref{gisource}) to find massless propagations in  general
relativity. Using \be \bigtriangleup \to -{\bf k}\cdot {\bf
k}=-k^2,~\ddot{t}_{ij} \to -\varpi^2 t_{ij}, \ee propagators with
source are derived as~\cite{Rub}\bea \label{einprp1}&&\ddot{t}_{ij}-
\bigtriangleup t_{ij}=-T_{ij} \to
t_{ij}(k)=\frac{T_{ij}}{\varpi^2-k^2}, \\
\label{einprp2}&&w_i=\frac{T_{0i}}{\bigtriangleup} \to w_i(k)=-\frac{T_{0i}}{k^2},\\
\label{einprp3}&&\psi=\frac{T_{00}}{2\bigtriangleup} \to
\psi(k)=-\frac{T_{00}}{2k^2},\\
\label{einprp4}&&\Phi=\frac{1}{2\bigtriangleup}\Big[T_{ii}+T_{00}-3\frac{\ddot{T}_{00}}{\bigtriangleup}\Big]\to
\Phi(k)=-\frac{1}{2k^2}\Big[T_{ii}+T_{00}-\frac{3\varpi^2T_{00}}{k^2}\Big],
\eea
 where $\Phi=A-\dot{\Pi}$ is a gauge-invariant scalar which plays the role of  Newtonian potential.
The above  shows clearly that tensor modes $t_{ij}$ are propagating
on the Minkowski background, while vector and  all scalars are
non-propagating because there is no kinetic terms $\varpi^2$ (second
order temporal derivative terms). It confirms  that the $\lambda=1
R$-model  has two propagating degrees of freedom for a massless
graviton, which is equivalent to the general relativity. All
propagators (\ref{einprp1})-(\ref{einprp4}) are those of  the dHL
gravity in the limit of $\omega \to \infty$ and in the massless
limit.

\subsection{Tensor modes}
The tensor equation takes a relatively simple form \be
\label{tensors} \ddot{t}_{ij}-c^2 \bigtriangleup t_{ij} +c^2m^2_2
t_{ij}+\frac{2c^2}{\omega}\bigtriangleup^2t_{ij}-\frac{\kappa^4\mu}{4\eta^2}\epsilon_{ilm}\partial^l\bigtriangleup^2t_j~^m-
\frac{\kappa^4}{4\eta^4} \bigtriangleup^3 t_{ij}=-T_{ij}. \ee We
find that there is no the vDVZ discontinuity because a single mass
term $m_2^2(\ge 0)$ is present. It is obvious that for $T_{ij}=0$,
the above equation reduces to Eq.(\ref{tensormo}). In the massless
limit of $m_2^2 \to 0$, Eq.(\ref{tensors}) has described chiral
gravitational waves without ghost propagating on the Minkowski
background~\cite{Myung4}. In the massive case, (\ref{tensors}) may
describe massive chiral gravitational waves.

\subsection{Vector modes}
 From
(\ref{vector1}) and (\ref{gisource}), the vector equations are
derived as  \bea \label{vectors1} &&\bigtriangleup
(V_i-\dot{F}_i)-m_1^2 V_i=T_{0i}, \\
\label{vectors2}&&\dot{V}_i-\ddot{F}_i-\tilde{m}_2
F_i=\frac{1}{\bigtriangleup} \dot{T}_{0i}. \eea
 From  equation (\ref{vectors1}), we find \be
V_i=\frac{\bigtriangleup}{\bigtriangleup-m_1^2}
\dot{F}_i+\frac{1}{\bigtriangleup-m_1^2}T_{0i}. \ee Plugging this
into equation (\ref{vectors2}) leads to \be
F_i=-\frac{1}{\bigtriangleup\Big(\partial_t^2-c^2\frac{m_2^2}{m_1^2}\bigtriangleup+c^2m_2^2\Big)}\dot{T}_{0i}
=-\frac{1}{c^2\bigtriangleup\Big(\partial_0^2-\frac{m_2^2}{m_1^2}\bigtriangleup+m_2^2\Big)}\dot{T}_{0i}.\ee
We define  the massless limit of the Lorentz-violating mass term
as~\cite{Rub}
 \be \label{masslessl} {\rm MLLV}:~~ m_i^2
\to 0,~~\frac{m_i^2}{m_j^2} \to {\rm fixed},~~ i,j=1,\cdots ,4,
\ee while the massless limit of the Lorentz-invariant Fierz-Pauli
mass term is defined by \be \label{masslessFP} {\rm MLFP}:~~
m_i^2=m^2 \to 0,~~ i,j=1,\cdots ,4. \ee
 The gauge-invariant vector takes the form \bea
w_i&=&V_i-\dot{F}_i=\frac{m_1^2}{\bigtriangleup-m_1^2} \dot{F}_i+
\frac{1}{\bigtriangleup-m_1^2}T_{0i}, \nn \\
&=&-\frac{m_1^2}{\bigtriangleup-m_1^2}\frac{1}{c^2\bigtriangleup
\Big(\partial_0^2-\frac{m_2^2}{m_1^2}\bigtriangleup+m_2^2\Big)}\ddot{T}_{0i}+
\frac{1}{\bigtriangleup-m_1^2}T_{0i}.
 \eea
Under the MLLV of (\ref{masslessl}), the gauge-invariant vector
reduces to (\ref{einprp2}) \be \label{vecfe}
w_i=\frac{1}{\bigtriangleup}T_{0i}, \ee which shows that the
vector is non-propagating in the massless limit. Also, we note
that nothing changes for the Fierz-Pauli case of $m_1^2=m_2^2$
because the gauge-invariant vector takes the form \be
w_i=-\frac{m^2}{\bigtriangleup-m^2}\frac{1}{c^2\bigtriangleup
\Big(\partial_0^2-\bigtriangleup+m^2\Big)}\ddot{T}_{0i}+
\frac{1}{\bigtriangleup-m^2}T_{0i}, \ee which leads again to
Eq.(\ref{vecfe}) in the MLFP of Eq.(\ref{masslessFP}).

\subsection{Scalar modes with $\lambda=1$}
In the scalar sector,   the field equations are obtained by
variation of (\ref{scalarL})+(\ref{gisource}) with respect to
$A,B,E,$ and $\psi$ as

\bea \label{ss1} && 2\bigtriangleup \psi-
\frac{\tilde{m}_4^2}{c^2}(\bigtriangleup E+3\psi )=\frac{T_{00}}{c^2}, \\
\label{ss2} && 2 \dot{\psi}-m_1^2 B=\frac{1}{\bigtriangleup}\frac{\dot{T}_{00}}{a}, \\
\label{ss3} &&2 \ddot{\psi}-\tilde{m}_2^2(\bigtriangleup E+\psi)
+\tilde{m}_3^2 ( \bigtriangleup E +3 \psi) -\tilde{m}_4^2
A=\frac{1}{\bigtriangleup }\frac{\ddot{T}_{00}}{a}, \\
\label{ss4} && 2\bigtriangleup \Phi-2c^2 \bigtriangleup \psi
+2\tilde{m}_2^2 \bigtriangleup
E+\frac{\kappa^4\mu^2}{8}\bigtriangleup^2\psi=t_{ii}-\frac{3}{\bigtriangleup}
\frac{\ddot{T}_{00}}{a}, \eea where $\Phi=c^2A-2\dot{B}+\ddot{E}$ is
the Newtonian  potential~\cite{Rub,RT}. Eq.(\ref{ss1}) provides \be
E=\frac{2c^2}{\tilde{m}_4^2} \psi-\frac{3}{\bigtriangleup}\psi
-\frac{T_{00}}{\tilde{m}_4^2 \bigtriangleup}, \ee while
Eq.(\ref{ss2}) gives \be B=
\frac{2\dot{\psi}}{m_1^2}-\frac{1}{m_1^2\bigtriangleup}
\frac{\dot{T}_{00}}{a}. \ee Eq.(\ref{ss3}) leads to \be
A=\frac{1}{\tilde{m}_4^2} \Bigg(2
\ddot{\psi}-\frac{2c^2(\tilde{m}_2^2-\tilde{m}_3^2)}{\tilde{m}_4^2}
\bigtriangleup \psi +2\tilde{m}_2^2 \psi+
\frac{\tilde{m}_2^2-\tilde{m}_3^2}{\tilde{m}_4^2}T_{00}-\frac{\ddot{T}_{00}}{a\bigtriangleup}\Bigg).
\ee Substituting these expressions into Eq.(\ref{ss4}) together
with $a=c^2$, one finds \bea &&
-6\ddot{\psi}+8\Big(\frac{1}{m_4^2}-\frac{1}{m_1^2}\Big)\bigtriangleup
\ddot{\psi}\nn \\
&&+c^2\Bigg[\Big(8\frac{m^2_2}{m_4^2}-2\Big)\bigtriangleup \psi
-4\Big(\frac{m_2^2-m_3^2}{m_4^4}\Big) \bigtriangleup^2 \psi
 -6m_2^2 \psi \Bigg] \nn \\
 &&=4\Big(\frac{1}{m_4^2}-\frac{1}{m_1^2}\Big)\frac{\ddot{T}_{00}}{c^2}-
 2\Big(\frac{m_2^2-m_3^2}{m_4^4}\Big)\bigtriangleup
T_{00} -\frac{3}{\bigtriangleup} \frac{\ddot{T}_{00}}{c^2}+ T_{ii}+
2\frac{m_2^2}{m_4^2} T_{00} \label{psifeq}. \eea Once that time
derivatives of $\psi$ and its  source $T_{00}$ are neglected, the
above equation could be solved to give  \be
\psi=\frac{n_1\bigtriangleup +n_0}{d_2\bigtriangleup^2+
d_1\bigtriangleup+d_0},\ee where the $n_i$ and $d_i$ are polynomials
in the masses.

The physics relevant to the vDVZ discontinuity is captured by
expanding $\psi$ in powers of $\frac{1}{\bigtriangleup}$ (that is,
$\bigtriangleup \gg m_i^2$) as  \be \psi=\frac{T_{00}}{2
\bigtriangleup}+\frac{c_1(m_i^2)}
{\bigtriangleup^2}+\frac{c_2(m_i^2)}{\bigtriangleup^3} \ee in the
$c=1$ units.  In the MLLV of (\ref{masslessl}) with
$c_1(m_i^2)=c_2(m_i^2)=0$, we obtain the same form as in
(\ref{einprp3}) \be \label{GRF}
\psi=\frac{1}{2\bigtriangleup}T_{00},\ee which implies that there is
no discontinuity at small distances. Also we find that $E$ and $B$
are finite in the massless limit. Therefore, there is no vDVZ
discontinuity in the scalar sector.

However, for the Fierz-Pauli case, we find from (\ref{psifeq}) and
(\ref{masslessFP}) that $\psi$ leads to \be
\psi=\frac{T_{ii}+2T_{00}-\frac{3\ddot{T}_{00}}{c^2\bigtriangleup}}{6c^2\square},\ee
which takes a further  form in the $c=1$ units \be \label{fpt}
\psi=\frac{T_{00}}{2\bigtriangleup}+\frac{T_{ii}-T_{00}}{6\square}.
\ee This confirms the presence of the vDVZ discontinuity of the
Fierz-Pauli case, as in Einstein gravity because the last term
implies that $\psi$ is a  propagating degree of freedom.
Consequently, we have shown that without the projectability
condition, the Ho\v{r}ava scalar $\psi$ is related to a scalar
degree of freedom appeared in the massless limit of a massive
graviton~\cite{CNPS}.

\section{Discussions}

We have   studied   graviton propagations of  scalar, vector,
  and tensor modes in the dHL  gravity ($\lambda R$-model) without projectability condition.
It is emphasized that   the quadratic Lagrangian is invariant under
diffeomorphism only for $\lambda=1$ case. This contradicts to the
fact that $\lambda$ is irrelevant to a consistent Hamiltonian
approach to the $\lambda R$-model~\cite{BR}. As far as scalar
propagations are concerned,
  there is no essential difference between dHL gravity ($\lambda R$-model) and general relativity.
    This implies that there are two degrees of freedom for a massless graviton without Ho\v{r}ava scalar,
    and five degrees of freedom including  Ho\v{r}ava scalar  appear for a massive graviton when  introducing
Lorentz-violating  and Fierz-Pauli mass terms. Importantly, the
strong coupling problem is not serious for vector and scalar modes
when choosing Lorentz-violating mass term, as was claimed in general
relativity~\cite{Rub}.
 It is shown  that
for $\lambda=1$, the vDVZ discontinuity is absent in the massless
limit of Lorentz-violating mass terms by considering external source
terms.  The dHL  with $\lambda=1$ recovers nicely the general
relativity with (without) mass term  in the linearized level. At
this stage, we wish to distinguish the massless limit of massive dHL
gravity from the $\lambda \to 1$ limit of dHL gravity. The former
case provides one scalar ($\psi$) propagation as was shown in
(\ref{fpt}) when choosing the Fierz-Pauli mass term, while it
provides no scalar propagation as was shown in (\ref{GRF}) when
choosing the Lorentz-violating mass term. The latter provides no
scalar propagation,  as was shown in (\ref{einprp3}).

On the other hand, the other case was  the dHL gravity  with the
projectability condition, the $S^{\lambda R}_2$ is invariant under
FDiff transformation for generic $\lambda$. In this case, the lapse
perturbation $A(t)$ is not a Lagrange multiplier but a
parameter~\cite{Myungm}. The gauge transformations for modes is the
same as in (\ref{trans1}) and (\ref{trans2}) except $ B \to
\tilde{B}=B+\dot{\xi}$ and thus, gauge-invariant scalars are $\psi$
and $\Pi=2B-\dot{E}$. However, this case gives rise to some
difficulty in performing a consistent Hamiltonian analysis because
the lapse perturbation $A$ plays no role. For $1/3<\lambda <1$, the
Ho\v{r}ava scalar $\psi$ suffers from  the ghost instability. Adding
the Lorentz-violating mass term did not cure the ghost
instability~\cite{Myungm}. In order to avoid the ghost instability,
one requires that the sound speed square $c^2_\psi$ be negative,
leading to the gradient instability for $\lambda>1$. To resolve this
gradient instability, one has to impose the limit of $\lambda \to
1$, which leads to the strong coupling problem~\cite{BPS,KA}.
However, it was suggested that there are many ways to tame the
gradient instability of Ho\v{r}ava scalar~\cite{IM}.

Finally, we ask whether  the projectability condition is really
essential for being unable to rescue  the dHL gravity from a
doubtable modified gravity seriously. If one abandons an original
constructing principle of the projectability inspired by condensed
matter physics, one has likely found   general relativity in the IR
limit.  However, we wish to point out that  although the $\lambda
R$-model has contributed to making  a progress toward a consistent
Hamiltonian approach to the dHL gravity,  there remains a subtle
issue on the equivalence between a  gauge-fixed version of general
relativity ($\lambda R$-model)  and general relativity (see footnote
1).

\section*{Acknowledgement}

This work  was supported by Basic Science Research Program through
the National  Research  Foundation (NRF)  of  Korea funded by the
Ministry of Education, Science and Technology (2009-0086861).

\end{document}